\documentclass[seceq]{ptptex}

\usepackage{graphicx}
\title{%        %You can use \\ for explicit line-break.
Hybrid neutron stars based on a modified PNJL model%
}

\author{%       %Use \scshape for the family name.
David \textsc{Blaschke}$^{a,b}$, %
Jens \textsc{Berdermann}$^{c}$
and 
Rafa{\l} \textsc{{\L}astowiecki}$^{a}$
}

\inst{%     %Affiliation, neglected when [addenda] or [errata].
$^{a}$Institute for Theoretical Physics, 
University of Wroc{\l}aw, 50-204 Wroc{\l}aw, Poland\\
$^{b}$Bogoliubov Laboratory for Theoretical Physics, JINR, 141980 Dubna, 
Russia\\
$^{c}$DESY, Platanenallee 6, D-15738 Zeuthen, Germany 
}

\abst{%         %This abstract is neglected when [addenda] or [errata].
We discuss a three-flavor Nambu--Jona-Lasinio (NJL) type quantum field 
theoretical approach to the quark matter equation of state (EoS)
with scalar diquark condensate, isoscalar vector mean field and 
Kobayashi-Maskawa-'t Hooft (KMT) determinant interaction.
While often the diquark and vector meson couplings are considered as free
parameters, we will fix them here to their values according to the Fierz 
transformation of a one-gluon exchange interaction. 
In order to estimate the effect of a possible change in the vacuum pressure
of the gluon sector at finite baryon density
we exploit a recent modification of the Polyakov-loop NJL (mPNJL) model
which introduces a parametric density dependence of the Polyakov-loop 
potential also at $T=0$, thus being relevant for compact star physics. 
We use a Dirac-Brueckner-Hartree-Fock (DBHF) EoS for hadronic matter phase 
and discuss results for mass-radius relationships following from a solution of 
the TOV equations for such a hybrid EoS in the context of observational 
constraints from selected objects.}
\begin{document}

\maketitle

\section{Introduction}
In the present contribution we will introduce a model for the EoS of
hybrid stars with deconfined quark matter cores and recent observational 
constraints \cite{Lattimer:2006xb,Klahn:2006ir} for their masses ($M$) 
and radii ($R$).

The status of the theoretical approach to the neutron star matter equation of
state  is very different from that for the high-temperature case at low or 
vanishing net baryon densities, where {\it ab initio} lattice QCD simulations 
provide EoS with almost physical quark masses systematically approaching the 
continuum limit \cite{Bazavov:2009zn}.
This guidance is yet absent at zero temperature and high baryon number 
densities, where a variety of models exists on different levels of the 
microphysical description which make different predictions for the state 
of matter.  
It is the current hope that the situation might change in a not too far future
when very accurate measurements of the $M-R$ relationship for compact stars 
become possible, e.g., with the {\it International X-Ray Observatory (IXO)}
project \cite{Paerels:2009pz}.
Once we are in possession of such data which allow upon inversion of the 
Tolman-\-Oppenheimer-\-Volkoff (TOV) equations the extraction of the cold dense
compact star EoS \cite{Steiner:2010fz},
we will face the problem of interpreting the physical 
content of this numerical result. 
Then, a broad basis of alternative EoS models like the one presented here
may become very useful.

A common feature of present hybrid star models is that the transition from 
hadronic to quark matter cannot yet be described on a unique footing where 
hadrons would appear as bound states
(clusters) of quarks and their possible dissociation at high densities 
as a kind of Mott transition \cite{Mott:1968zz} like in nonideal plasmas 
\cite{Redmer:1997} or in nuclear matter \cite{Ropke:1982,Typel:2009sy}. 
Early nonrelativistic potential model approaches 
\cite{Horowitz:1985tx,Ropke:1986qs} are insufficient since they 
cannot accomodate the chiral symmetry restoration transition in a proper way.
Therefore, at present the discussion is restricted to so-called two-phase 
aproaches where the hadronic and the quark matter EoS are modeled separately
followed by a subsequent phase transition construction to obtain a hybrid EoS.

Widely employed for a description of quark matter in compact stars are 
thermodynamical bag models of three-flavor quark matter with 
eventually even density-dependent bag pressure $B(n)$, as in Ref. 
\cite{Baldo:2003vx}. 
A qualitatively new feature of the phase structure appears
in chiral quark models of the Nambu--Jona-Lasinio type which describe the 
dynamical chiral symmetry breaking of the QCD vacuum and its partial 
restoration in hot and dense matter, see Ref. \cite{Buballa:2003qv} for a 
review. In contrast to bag models, in these approaches at low temperatures 
the light and strange quark degrees of freedom may appear sequentially with 
increasing density \cite{Gocke:2001ri,Blaschke:2008vh,Blaschke:2008br}, 
so that strangeness may even not 
appear in the quark matter cores of hybrid stars, before their maximum mass
is reached. 
Once chiral symmetry is restored, a rich spectrum of color superconducting 
quark matter phases may be realized in dense quark matter, depending on it's 
flavor composition and isospin asymmetry  \cite{Alford:2007xm} with 
far-reaching consequences for hybrid star phenomenology, e.g., $M-R$ 
relationships and cooling behavior. 

We will consider here a color superconducting three-flavor NJL model with 
selfconsistently determined density dependences of quark masses and scalar 
diquark gaps, developed in Refs.  
\cite{Ruester:2005jc,Blaschke:2005uj,Abuki:2005ms}, 
including the flavor-mixing KMT determinant interaction 
\cite{Kobayashi:1970ji,'tHooft:1976up}.   
Only recently it became clear \cite{Agrawal:2010er,Blaschke:2010vd} 
that this flavor mixing is crucial for the possible stability of strange 
quark matter phases in hybrid stars.
As a new aspect, we will couple the chiral quark sector to a Polyakov-loop
potential which is a modification \cite{Dexheimer:2009va} of the one used 
in standard PNJL models \cite{Ratti:2005jh,Roessner:2006xn}.
 
\section{Modified PNJL model with color superconductivity}

We start from the path integral representation of the partition function for 
the modified color superconducting three-flavor PNJL model 
\begin{eqnarray}
\label{partition}
	Z[T,V,\{\mu\}]=\int{\mathcal D}\bar{q}{\mathcal D}q~
{\rm e}^{-\int^\beta d^4x \{\bar{q}[i\gamma^\mu\partial_\mu - \hat{m} 
-\gamma^0 (\hat{\mu}+i\lambda_3\phi_3)]q 
-{\mathcal L}_{\rm int}+{\mathcal U}(\Phi)\}}
\end{eqnarray}
with the interaction Lagrangian 
${\mathcal L}_{\rm int}={\cal L}_{\bar{q}q}+{\cal L}_{qq}$, where 
\begin{eqnarray}
	{\cal L}_{\bar{q}q} &=& G_S \sum_{a=0}^8 \Big[({\bar q}\tau_a q)^2
	+ ({\bar q} i\gamma_5\tau_a q)^2\Big] 
	+ G_V({\bar q} i\gamma_0 q)^2
\nonumber \\
	&-&K\left[{\rm det}_f(\bar{q}(1+\gamma_5)q)
	+{\rm det}_f(\bar{q}(1-\gamma_5)q)\right],
\label{qbarqlag} \\
	{\cal L}_{qq} &=& G_D\!\!\!\!\!\!\sum_{a,b = 2,5,7}\!\!\!\!\!\! 
%\sum_{b = 2,5,7}
	(\bar{q} i\gamma_5 \tau_a \lambda_b C\bar{q}^T)
	(q^T C i\gamma_5\tau_a\lambda_b\,q),
\label{qqlag}
\end{eqnarray}
with $\tau_a$ and $\lambda_b$ being the antisymmetric Gell-Mann matrices 
acting in flavor and color space, respectively.
We have suppressed the flavor index for the current quark mass matrix 
$\hat{m}$ and the chemical potential matrix $\hat{\mu}$. 
The scalar ($G_S$), diquark ($G_D$), vector ($G_V$) and KMT ($K$) couplings are
to be determined by hadron phenomenology, see \cite{Rehberg:1995kh}. 
The Polyakov-loop $\Phi={\rm Tr}_c[\exp(i\beta\lambda_3\phi_3)]/N_c$ is an 
order parameter for confinement, weighted with the phenomenological 
potential 
\begin{eqnarray}
\label{PL}
{\mathcal U}(\Phi)=(aT^4+b\mu^2T^2+c\mu^4)\Phi^2 
+a_2T_0^4 \ln(1-6\Phi^2+8\Phi^3-3\Phi^4)~,
\end{eqnarray}
which is a modification \cite{Dexheimer:2009va} of the standard PNJL model,
now accounting for an explicit chemical potential dependence of 
${\mathcal U}(\Phi)$, even at $T=0$, which is not present in the traditional 
parametrizations by, e.g., Refs. \cite{Ratti:2005jh,Roessner:2006xn}.  

Our description of quark matter is based on the grand canonical thermodynamic 
potential 
~\cite{Buballa:2003qv,Ruester:2005jc,Blaschke:2005uj,Abuki:2005ms,Klahn:2006iw}
which in the meanfield (stationary phase) 
approximation to the partition function (\ref{partition}) is given by 
\begin{eqnarray}
	\Omega_{\rm MF}(T,\{\mu\}) &=& \frac{\phi^2_u+\phi^2_d+\phi^2_s}{8 G_S}
	+\frac{K\phi_u\phi_d\phi_s}{16G^3_S} 
	-\frac{\omega^2_u+\omega^2_d+\omega^2_s}{8 G_V}
%\nonumber \\
	+\frac{\Delta^2_{ud}+\Delta^2_{us}+\Delta^2_{ds}}{4 G_D} 
\nonumber \\
	&-&\int\frac{d^3p}{(2\pi)^3}\sum_{n=1}^{18}
	\left[E_n+2T\ln\left(1+e^{-E_n/T}\right)\right] 
%\nonumber \\
%	+ \Omega_{lep} 
	- \Omega_0 + {\mathcal U}(\Phi_0)~,
\label{potential}
\end{eqnarray}
where $E_n=E_n(p,\,\mu;\,\mu_Q,\mu_3,\mu_8,\,\phi_u,\phi_d,\phi_s,\,
\omega_u,\omega_d,\omega_s,\,\Delta_{ud},\Delta_{us},\Delta_{ds},\Phi_0)$ 
are the quasiparticle dispersion relations, 
obtained by numerical diagonalization of the quark propagator matrix in 
color-, flavor-, Dirac- and Nambu-Gorkov spaces.
The values of the meanfields (order parameters) are obtained from a 
minimization of $\Omega(T,\{\mu\})$, which is equivalent to the selfconsistent 
solution of the set of corresponding gap equations. 
The subtraction of $\Omega_0$ garantees that in the vacuum $\Omega(0,\,0)=0$. 
For applications to compact stars, we will have to add the contribution from
leptons ({\it e.g.}, electrons, muons and the corresponding neutrino flavors) 
and obey the constraints of color and electric neutrality as well as 
$\beta$-equilibrium of weak interactions between quark flavors and leptons. 

The mean-field contribution of the Polyakov-loop potential could be viewed
as a $T,~\mu$-dependent modification of the bag function $\Delta B(T,\mu)={\mathcal U}(\Phi_0)$ 
which accounts for possible changes in the pressure of the gluon sector
related to a partial melting of the gluon condensate at finite $T$ and $\mu$.
Therefore, it takes negative values.
In the present contribution, we will discuss exploratory calculations of those 
effects for compact star structure, we will compare results with the above 
modified PNJL (mPNJL) model with those of approximating the influence of these
effects by a simple (negative) bag constant $\Delta B$.
Note that the mPNJL model provides us with a more micoscopic picture on the
origin of density (and temperature) dependence of nonperturbative QCD 
thermodynamics which in some phenomenological approaches are subsumed in a 
density dependent bag constant as, e.g., in Ref.~\cite{Baldo:2003vx} 

As pointed out in \cite{Buballa:2003qv}, due to the mixing of the light and 
strange flavor sectors by the KMT interaction, the difference in the critical 
chemical potentials for the chiral phase transitions in these sectors 
(which coincide with the onset of 2SC and CFL phases, respectively) gets 
diminished. 
This entails that the phase transition between hadronic matter 
(described by a realistic nuclear EoS, e.g., the DBHF one, see 
\cite{Klahn:2006iw}) and superconducting quark matter may eventually proceed 
directly into the CFL phase, provided the diquark coupling is sufficiently 
strong, see Ref.~\cite{Blaschke:2010vd}.

In the present study, we will not investigate the dependence of the 
thermodynamics and phase structure on the strengths of the coupling constants
which was done in a previous work \cite{Blaschke:2010vd}, 
but rather adopt for their ratios with the scalar coupling strength the 
values given by the Fierz transformation of the one-gluon exchange interaction,
i.e., $G_V=0.5~G_S$ and $G_D=0.75~G_S$. The values of $G_S$, cutoff $\Lambda$ 
and current quark mases $m_{u,d}$ and $m_s$ are chosen such as to describe the 
pion decay constant, pion and kaon masses as well as a light quark constituent 
mass of 350 MeV. The KMT coupling is fixed such that the $\eta-\eta'$ mass 
splitting is obtained. 

\begin{figure} [!ht] 
\includegraphics[width=0.7\textwidth,height=0.4\textwidth]{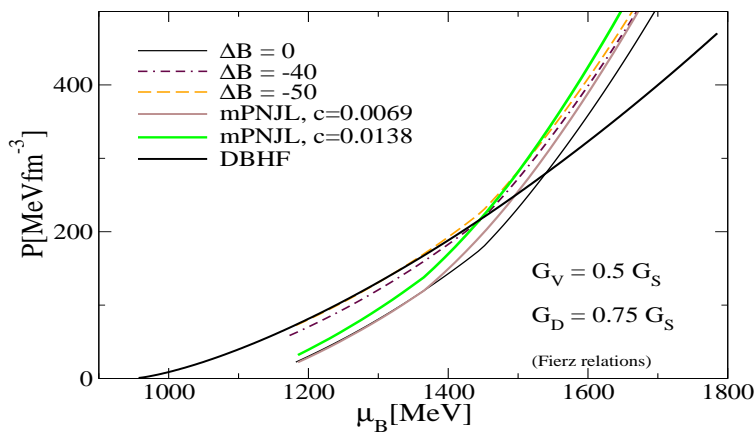}
\caption{ Equations of state for neutron star matter in beta-equilibrium. 
The hadronic phase is described by the DBHF EoS (solid line) and the 
quark matter EoS correspond to the mPNJL model (\ref{potential}) and its 
abridged version, for details see text. 
}
\label{f:1} 
\end{figure}

In Fig.~\ref{f:1} we show the $T=0$ EoS as $P(\mu)$ for the {\it ab initio} 
DBHF approach to nuclear matter together with the results of evaluating the 
mPNJL model for quark matter in meanfield approximation (\ref{potential})
for specific choices of the coefficient $c$ in ${\mathcal U}(\Phi_0)$. 
It is interesting to note that those results compare well with those of the
simplifying ansatz ${\mathcal U}(\Phi_0)={\mathcal U}(\Phi_0;\mu=0)+\Delta B$, 
for the region close to the hadron-to-quark matter transition as relevant
for applications to compact star structure. 
Note, that in the present model the range of $\Delta B$-values is limited to 
$|\Delta B|<50$ MeV/fm$^3$. For the mPNJL model, there is no such 
limitation of the values for the parameter $c$.

\begin{figure} [!ht] 
\includegraphics[width=0.9\textwidth,height=0.4\textwidth]{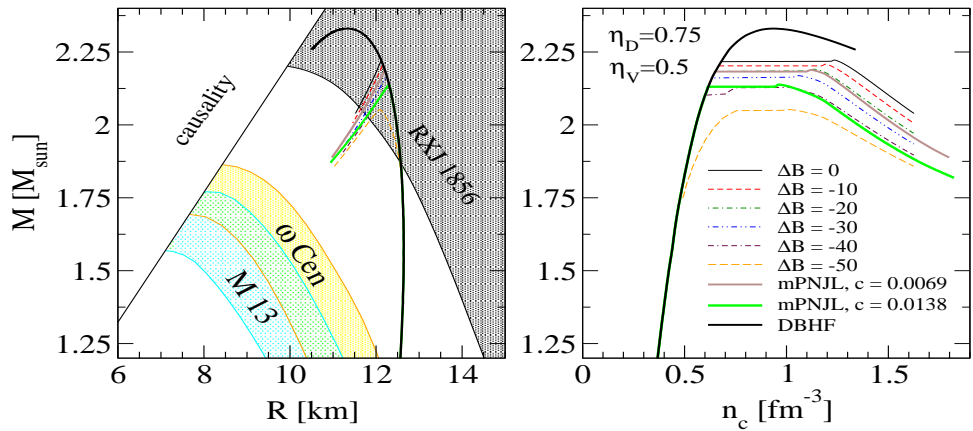}
\caption{Compact star sequences for different values of the  $c$ parameter
(bag constant $\Delta B$ in MeV/fm$^3$) in the (abridged) mPNJL model 
(\ref{PL}). 
For a discussion, see text.
%For $B=-50$ MeV/fm$^3$ there is a sequence of stable hybrid stars with 
%2SC quark core,  in the mass range $1.9<M/M_\odot<2.04$ before the hybrid 
%stars turn unstable at the onset of the CFL phase in the quark matter core.
%For the other parameter choices, there is a direct transition from hadronic 
%to CFL quark matter with a very small sequence of marginally stable CFL 
%quark core hybrid stars at the respective maximum mass. 
%For comparison recent mass radius constraints are shown coming from
%RX J1856 \cite{Trumper:2003we,Klahn:2006ir},  M13 \cite{Gendre:2003pw} and 
%$\omega$ Cen \cite{Gendre:2002rx}, see also \cite{Lattimer:2006xb}.
}
\label{f:2} 
\end{figure} 

In Fig.~\ref{f:2} we show the $M-R$ and $M-n_c$ relationships for nonrotating
compact star sequences obtained as solutions of the Tolman-Oppenheimer-Volkoff 
equations for the hybrid EoS shown in Fig.~\ref{f:1} where
the $c$-parameter of the mPNJL model (\ref{PL}) and the bag constant $\Delta B$
of the abridged mPNJL model are varied as respective free parameters. 
The diquark and the vector couplings are set to the values of the Fierz 
relation for one-gluon exchange, $\eta_V=G_V/G_S=0.5$ and $\eta_D=0.75$, resp. 
For $\Delta B$[MeV/fm$^3$] $ = -50$ and $-40$ we obtain sequences of 
stable hybrid stars with 2SC quark core, in the mass range 
$1.74<M/M_\odot<2.04$ and $2.10<M/M_\odot<2.12$, respectively.
For the other parameter choices, there is a direct transition from hadronic 
to CFL quark matter with a very small sequence of marginally stable CFL 
quark core hybrid stars at the respective maximum mass. 
A more detailed investigation is subject to ongoing work.

The question arises whether flavor-mixing interactions other than the KMT one
have to be invoked. 
It seems necessary in a next step to include channels which appear 
after Fierz transformation of the KMT interaction and couple 
chiral condensates with diquark condensates  \cite{Steiner:2005jm}.
This may lead to a new critical point in the QCD phase diagram 
\cite{Hatsuda:2006ps} and, depending on the sign of the coupling, to a further
reduction of the strange chiral condensate which enforces the flavor mixing
effect studied here.  

The observational constraints for masses and radii are not yet settled. There
is a lower limit for the $M-R$ relation from RXJ 1856.5-3754
\cite{Trumper:2003we} which requires either a large radius $R>14$ km for a 
star with $M=1.4~M_\odot$ or a mass larger than $\sim 2~M_\odot$ for a star 
with 12 km radius. 
The latter is accomodated with the present EoS.
$M-R$ relations for the quiescent binary neutron stars in globular clusters
M13 \cite{Gendre:2003pw} and $\omega $ Cen  \cite{Gendre:2002rx} point to 
rather light and compact stars as described by those sequences obtained here,
%However, as a caveat of the present model, it is difficult to obtain 
but without quark matter cores for 
%compact stars with typical masses or even with presently determined maximum 
masses $M<1.9~M_\odot$.  

In order to test any of the predictions for quark matter phases in compact 
stars and their possible consequences, we look forward to future observational 
campaigns devoted to determine the masses \cite{Freire:2009dr} and $M-R$ 
relationship for compact stars \cite{Ozel:2009da} with high precision and thus 
to constrain the dense matter EoS \cite{Steiner:2010fz}. 
 
\section{Conclusions}

The effect of density-dependent modifications of the gluon condensate
on the sequential occurrence of superconducting two- and three-flavor quark 
matter phase transitions in the EoS for cold dense matter has been studied 
in a modified PNJL model with color superconductivity and flavor-mixing KMT 
determinant interaction. 
Different from previous work, here we fix the coupling constants from 
hadron phenomenology and Fierz relations which leaves the unknown chemical
potential dependence of the modified Polyakov-loop potential as a free 
parameter of the present study.
It is found that for sufficiently large values of the parameter $c$ there 
can be a stable quark matter core in massive hybrid stars.

\section*{Acknowledgements}
DB and RL acknowledge the hospitality of the Yukawa Institute for Theoretical 
Physics Kyoto, partial support by the Yukawa International Program for
Quark-Hadron Sciences and discussions during the Workshop program 
``New Frontiers in QCD'' where this work has been started. 
This work has been supported in part by the Polish Ministry of Science and 
Higher Education  (MNiSW) under grant No. N N 202 2318 37 (DB, RL), by 
the Russian Fund for Basic research (RFBR) under grant No. 08-02-01003-a (DB),
and by CompStar, a research networking programme of the European Science 
Foundation.

%\appendix
%\section{First Appendix} %Empty argument \section{} yields `Appendix'. 
%
%\section{Second Appendix}

\end{document}